\documentclass[twocolumn]{aa}
\usepackage{graphicx}
\usepackage{txfonts}
\begin{document}

\title{Precessing jets from a moving source and bright X-ray filaments in 
galaxy clusters} 

\author{
  Mario Rodr\'\i guez-Mart\'\i nez\inst{1,3}, 
  Pablo F. Vel\'azquez\inst{2},
  Luc Binette\inst{3}
  and Alejandro C. Raga\inst{2}} 
  \offprints{P.F. Vel\'azquez}

\institute{Harvard-Smithsonian Center for Astrophysics
60 Garden Street
Cambridge MA, MS-83,02138. USA. \\
  \email{mario@head.cfa.harvard.edu}
   \and
   Instituto de Ciencias
Nucleares, Ciudad Universitaria, Apartado Postal 70-543, CP 04510, M\'exico
D.F., M\'exico.\\ 
\email{pablo,raga@nucleares.unam.mx}
    \and
Instituto de Astronom\'\i{}a, Universidad Nacional
  Aut\'onoma de M\'exico, Ciudad Universitaria, Apartado Postal 70--248, CP
  04510, M\'exico D.F., M\'exico \\
  \email{binette@astroscu.unam.mx}}

   \date{Received ; accepted}

\abstract{We present hydrodynamical calculations carried out 
with the 3D {\sc yguaz\'u-a} code of a precessing jet model, which
interacts with a plane parallel wind.
This scenario describes an extragalactic jet, in which the jet source
is in motion with respect to the surrounding intra-cluster medium.
From the numerical results, synthetic emission maps and spectra in X-ray band
were obtained. We compare these predictions with observations of
the radio jets emanating
from the radio-galaxy 4C 26.42 (in the Abell 1795 galaxy cluster).  We
find that the general morphology of the radio jets can be
described by a point-symmetric precessing jet system interacting with a
plane parallel wind (i.~e., the intra-cluster medium flowing past the
galaxy). We also find that our synthetic X-ray
emission maps reproduce the observed large scale structures (with
sizes of the order of tens of kpc).

}
\authorrunning{Rodr\'\i guez-Mart\'\i nez et al. }
\titlerunning{Precessing jets and bright X-ray filaments in galaxy clusters}
\maketitle

\section{Introduction}
\label{sec:intro}

Extragalactic jets basically show three 
types of morphological structures (observed in radio-continuum,
\cite{zaninetti})~: the linear
shape, the tail or ``C'' shape and the ``Z'' (or ``S'') shape. The
``C'' and ``Z'' shapes are generally interpreted as being caused by either the
relative motion of the radio galaxy with respect to the ambient
medium (``C-shape'') or by the precession of the jet direction
(``Z''), respectively.  A combination of both effects is probably
present in most objects but one of them is usually starkly dominant. The
orientation of the observer relative to the jet direction also plays
an important role.

Several authors have studied extragalactic jets by means of  numerical
simulations. We can mention the MHD model developed by 
Pietrini \& Torricelli-Ciamponi
(1992) and Torricelli-Ciamponi \& Pietrini (1993), where they analyze the
formation of structures along the jet axis. Cox, Gull \& Scheuer (1991) 
carried out 3D numerical simulations of precessing jets in order to explain 
the formation of hot-spots and the general morphology observed in extragalactic 
radio sources.  Ferrari (1998) gives  a thorough review about the modeling 
of extragalactic jets.
Interestingly, in the arena of stellar jets, the
``S-shape'' has likewise been observed and modeled by invoking
precession, such as in HH 34 (Masciadri et al. 2002, and references therein)
or in the proto-planetary nebula Hen 3-1475 (Vel\'azquez, Riera \& Raga 2004 and
references therein).

As examples of extragalactic jet with
``S'' shapes, we can mention the jet system from the radio source 
4C 26.42 (in the Abell 1795
galaxy cluster, van Breugel et al. 1984 and Ge \& Owen 1993), the radio galaxies 3C294 
(Crawford et al. 2003), 
3C 334 (Kellermann \& Owen 1988) and 4C29.47 (Condon \& Mitchell 1984).
These jets are immersed in an intra-cluster medium (ICM)
and their radio lobes can be asymmetric due to the motion
of the jet source. This fact was reported by
Hill et al. (1988) for the case of the radio source 4C 26.42, which is 
is in motion within the gravitational potential of the Abell 1795 cluster,
having a peculiar radial velocity of 365
km\,s$^{-1}$. However, Oegerle \& Hill (1994) later revised this measurement 
and favored a lower value of 150 km\,s$^{-1}$.

While the properties of extragalactic jets can be studied with
radio-continuum observations, the ICM in which these objects are propagating
can be detected in H$\alpha$ or X-ray emission.
Recently, a new generation of X-ray satellites (with the Advanced Camera for
Imaging and Spectroscopy of {\sc Chandra}, and XMM-Newton) have obtained
images and spectra with good spatial and spectral resolution (respectively)
of galaxy clusters containing
radio jets. 

X-ray and radio observations provide evidence of jet-ICM
or ISM interactions.
Zanni et al. (2003) and Reynolds, Heinz \& Begelman (2001) (Rizza et al. 2000)
 carried out 2D (axisymmetric) and
3D numerical simulations, respectively, in order to analyze this interaction 
and its
observational consequences. They report two main regimes for the jet-ICM
interaction. In the first one, the jet cocoon expands into the ICM
supersonically
giving a shell-like morphology in the X-ray emission which surrounds
the radio-jets. The second regime is characterized by a subsonic 
cocoon expansion. In this case, the X-ray emission is strong at the head of
the jet, while the sides of the cocoon practically disappear.
Zanni et al. (2003) found
that the main parameter determining the appropriate regime is the jet
kinetic power (or ram pressure).

The models listed in the last paragraph predict shell-like or partial shell-like morphology for
the X-ray emission from jet-ICM interaction. However, several authors have
 reported that the X-ray emission of some galaxy
clusters has bright filamentary structures embedded into diffuse emission
(Fabian et al. 2001 for the
Abell 1795 case; Crawford et al. 2003 for 3C294).

In this paper we explore the possibility that the jet-ambient gas
interaction would lead to a filamentary morphology in X-ray,  as a result of 
two
processes: the precession of the point-symmetric jets and the displacement 
of the central galaxy with respect to the intra-cluster medium. We
present numerical simulations carried out with the 3D {\sc yguaz\'u-a} code
(Raga et al. 2000) which include these two effects.

This work is organized in the following way. In section 2 we present
the model employed in our simulations, giving the initial conditions
and describing the assumptions that have been made. We also explain how
the derived X-ray emission is obtained from our numerical results. In Section
3, we present both the hydrodynamical simulations and the predicted
X-ray emission maps. In Section 4 we analyze our numerical results and
compare with observations and finally in section 5 we give our final 
discussion.

\section{Description of the model}
\label{sec:model}

In order to model the interaction of an extragalactic radio-jet
with the surrounding environment, we consider a model
consisting of a point-symmetric, precessing jet system interacting with a
plane-parallel wind.

We consider a point-symmetric, precessing jet system of fixed jet velocity, 
where the precession is characterized by a semi-opening angle $\alpha$ 
and a precession period $\tau_{p}$.  The adopted parameters
are listed in
Table 1.  An important aspect to consider is that, for example, the ``S''
morphology of some 
radio-jets exhibits some degree of asymmetry between
the two radio lobes. This asymmetry can be produced by the
motion of the cD galaxy with respect to the surrounding intra-cluster
medium. Velocities in the range [150,
350]~km s$^{-1}$ have been observationally reported by Oergerle \& Hill (1994)
and Hill et al. (1988), for the case of the 4C 26.42 radio-jet.
This effect has been explicitly taken into account in our models
although, for computational purposes, the galaxy is deemed stationary
within the 3D grid and it is the intra-cluster medium which is moving
at a velocity of 300~km s$^{-1}$.

\subsection{Initial conditions and assumptions}
\label{sec:conditions}
In order to determine physical parameters of the extragalactic jets and the
ICM, we have considered the
observational results of van Breugel et al.(1984), Fabian
et al.(2001) and Ettori et al.(2002).
 
3D numerical simulations were carried out with  the  
{\sc yguaz\'u-a}   code (Raga, Navarro-Gonz\'alez \& Villagr\'an-Muniz 2000),
 which integrates the gas-dynamical equations with a  
second order accurate technique (in time and space) employing 
the flux-vector splitting method of van Leer (1982). 
Our numerical simulations
use a 5-level, binary adaptive grid, with a maximum resolution of 
5.86$\times$10$^{20}$~cm, in a computational domain of 
(1.5, 1.5, 3.0)$\times$10$^{23}$~cm, along the $x-$, $y-$ and $z-$ axis, 
respectively. The jets are injected at $x=y=7.5\times 10^{22}$~cm and 
$z=1.5\times 10^{23}$~cm (i.e. at the centre of the computational domain).
An injection jet velocity $v_j=10^{9}$~cm s$^{-1}$ 
a precession angle $\alpha=20\degr$, and a precession period 
$\tau_p=10^{15}$~s were considered. The jet precession cone axis
is aligned with the $z-$~axis.  

For the intra-cluster medium the temperature and the density were
fixed at $6\times 10^{7}$~K and 0.1~cm$^{-3}$ ($2.16 \times 10^{-25}$
g cm), respectively (based on {\sc Chandra} X-ray observations, Ettori et al.
2002). Also, the intra-cluster wind has a velocity of $3\times 10^7$~cm
s$^{-1}$, forming an angle of 60$\degr$ with respect to the
$x-$axis.  For the jets we adopted a temperature and density of
$6\times 10^{7}$~K and $5\times 10^{-3}$~cm$^{-3}$ ($1.1 \times
10^{-26}$ g cm$^{-3}$), respectively. The initial diameter and length of the
jet were of $6\times 10^{21}$~cm  (or 10 pixels)  and 
$4\times 10^{21}$~cm  (or
7 pixels), respectively. The jet and environmental parameters are 
summarized in Table 1.

The version of the {\sc yguaz\'u-a} code that we have used does not 
include radiative losses. This simplification is certainly acceptable
since we find that the typical dynamical time $\tau_d=4.7$~Myr
which characterizes our calculations is much shorter than the cooling 
time scale of 127 Myr.

\begin{table}
  \caption{Input parameters of model I}
  \begin{tabular}{ll}
            \hline
            \noalign{\smallskip}
   Parameter & Value\\
            \hline
            \noalign{\smallskip}
    $x$ size    & $1.5\times 10^{23}$~cm\\
    $y$ size    & $1.5\times 10^{23}$~cm\\
    $z$ size    & $3.0\times 10^{23}$~cm\\
    ambient density, $n_w$& 0.1~cm$^{-3}$ \\
    ambient temperature, $T_w$ & $6\times 10^7$~K \\
    wind velocity, $v_w$ & $3\times 10^7$~cm s$^{-1}$ \\
    wind entrance angle, $\beta$ & $60\degr$ \\
    jet density, $n_j$    & $5\times 10^{-3}$~cm$^{-3}$\\
    jet temperature, $T_j$ & $6\times 10^7$~K\\
    jet velocity, $v_j$& $10^9$~cm s$^{-1}$\\
    precession period, $\tau_p$ & $10^{15}$~s\\
    half precession angle, $\alpha$ & $20\degr$ \\
            \hline
            \noalign{\smallskip}
  \end{tabular}
\end{table}

Hereafter, we will call the model described in this subsection ``model I''.

We note that our jet has a Mach number of $\sim 20$, and at such a high
Mach number that the jet pressure does not play an important role. Also,
the ram pressure of the jet material is an order of magnitude higher
than the environmental pressure, so that the jet is not ``chocked'',
and propagates supersonically (see Figure 12 of Zanni et al. 2003).

\subsection{Other parameter values}
\label{sec:math}
In order to check how sensible our model I is with respect to changes of input
parameters, we  explored two other  models. In model II, we
have chosen a precession period five times shorter
($\tau_p=1.1\times 10^{14}$~s) than in model I, keeping constant the other
input parameters. 

In model III, we have instead changed  the incidence angle of the
plane-parallel wind. We set its value at 90$\degr$, i.e. the
plane-parallel wind is actually moving in the $-z$ direction.

\subsection{Simulated X-ray maps and spectra}
\label{genxray}

From our numerical results it is possible to obtain (through line-of-sight
integrations) X-ray emission maps and also X-ray spectra. In this way,
the numerical results can be directly compared with X-ray observations
from satellites such as {\sc Chandra}.  Such  comparisons have recently
been  made for other flows by Raga et al. (2001, for the case of the 
Arches cluster)
and Vel\'azquez et al. (2004, for the X-ray emission of the
supernova remnants W44 and G296.5+10.0).

To calculate the emission coefficient, $j_\nu (n,T)$, we have used the
{\sc chianti}~\footnote{The {\sc chianti} database and its associated
{\sc idl} procedures are freely available at:
{http://wwwsolar.nrl.navy.mil/chianti.html},
{http://www.arcetri.astro.it/science/chianti/chianti.html},
{http://www.damtp.cam.ac.uk/users/astro/chianti/chianti.html}} atomic
database and its associated {\sc idl} software (Dere et al. 2001,
Young et al. 2003).  We then generated the synthetic X-ray emission
coefficient (integrated over the energy range of interest) as a
function of temperature $T$ (at a reference density of 1~cm$^{-3}$) and
then scale it using the $j\propto n^2$ low density regime density
dependence.

We assumed that the gas is in coronal ionization equilibrium. For the
calculation of the X-ray emission we used the Mazzota et al. (1998)
coronal ionization equilibrium and the Anders \&
Grevesse (1989) solar abundances.

From our numerical models, we obtained synthetic emission maps corresponding
to the soft ([0.3--1.8]~keV) and hard ([1.8--7]~keV) energy bands
of the {\sc Chandra} satellite. For the simulated spectra, which
cover the whole energy band ([0.3--7]~keV), we considered the
effects of interstellar absorption using the absorption curve of
Morrison \& McCammon (1983) and integrated the emission
coefficient along either the $y-$axis ($xz$ projection) or the $x-$axis
($yz$ projection).

\section{Results}
\label{sec:res}
\subsection{Hydrodynamical results}
\label{sec:reshydro}
\begin{figure}[!t]
  \includegraphics[width=\columnwidth]{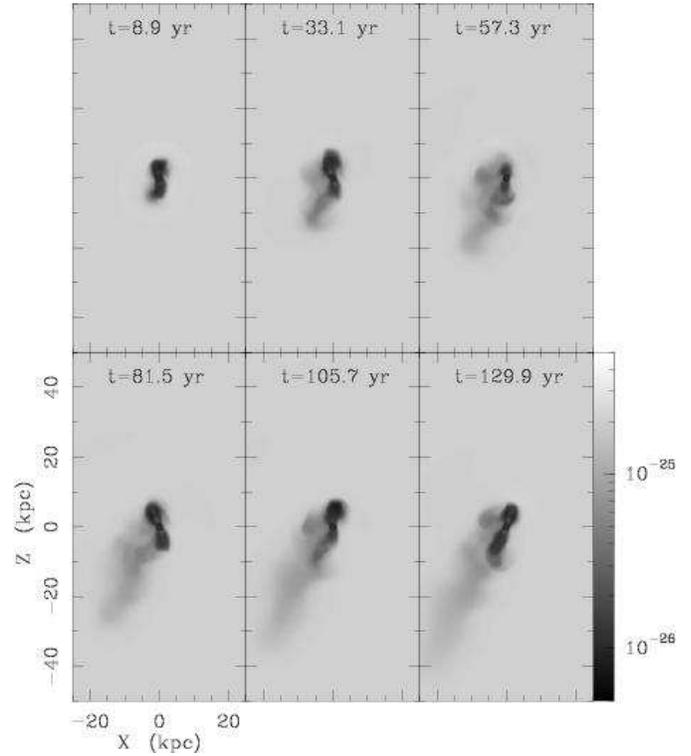}
  \caption{Time evolution of the density stratification in the $y=0$
  plane, for the jet-wind interacting model. Integration times ranges
  from 8.9 to 130 Myr. The logarithmic gray-scale for the density
  is given by the bar next to the bottom-right panel (in units of g
  cm$^{-3}$). The $x-$ and $z-$~axes are labeled in units of
  kpc. Note that darker areas correspond to lower density regions.} 
 \label{f1}
\end{figure}

For  analyzing the characteristics of the interacting jet-wind
model, the temporal evolution of the density distribution is shown in
Fig.\ref{f1}.  These maps are $y=0$ cuts and correspond
to the density distribution in the $xz$~plane. The density scale is
given in units of g~cm$^{-3}$. A time interval  $\Delta t=
24.2$~Myr separates the sequence of panels displayed in Fig.\ref{f1},
spanning a total temporal evolution of 130 Myr. A long wake can
be observed in Fig.\ref{f1}, as a result of the interaction between
the precessing jets and the plane-parallel ICM wind. This wake has
low-density gas, which is swept up by the ICM wind. The density of the
wake is of $\sim 5\times 10^{-26}$g cm$^{-3}$, i.e. five
times lower than its surrounding environment due to the mixing between the
shocked jet gas (by the Mach disk) and the more dense ICM gas, which has
been shocked by the jet bow shock. 
The jet axis  is continously varying its inclination 
with respect to the velocity of the
incident ICM wind. At smaller incident ICM wind-jet axis angles
the collision becomes stronger than in the  case with larger ICM wind-jet axis
angles (i.e. the ICM wind is actually hitting the cocoon side). 
This produces small differences in the density of the shocked ICM  gas, which
is given by:

\begin{equation}
{{\rho_2}\over{\rho_1}}={{(\gamma+1)M_1^2}\over{(\gamma-1)M_1^2 +2}}
\label{eqrho}
\end{equation}

\noindent where $\gamma$ is the specific heat ratio, $M_1$ is the shock 
Mach number, $\rho_2$ and $\rho_1$ are the shocked and unshocked ICM densities,
respectively. For strong shocks we have $M_1 >> 1$ and  
 the density ratio (see Eq.\ref{eqrho}) is  approximately equal to 4 (in the 
monoatomic gas case), while it is slightly smaller than this value, when the 
shocks  have  $M_1 \simeq 1$. Because of this and after the mixing with the
low density shocked jet gas, the wake has a 
knotty structure and its width is variable, when this shocked ICM-jet gas
  is swept up and pushed behind by the incident flow.

\begin{figure}[!t]
  \includegraphics[width=\columnwidth]{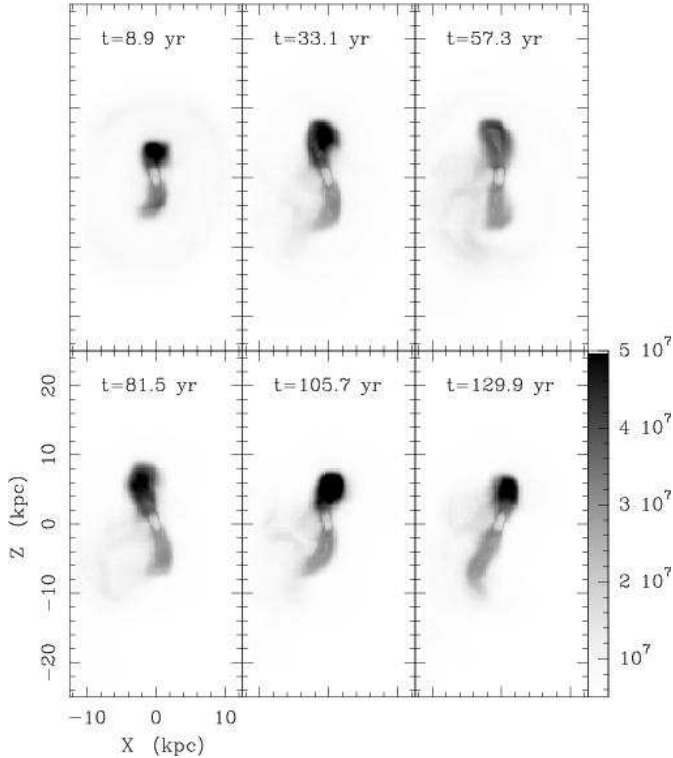}
  \caption{Temporal evolution of the temperature averaged along lines
  of sight. Each $xz$
  projection map has been obtained by integrating the temperature
  along the $y$ axis. High-temperature regions (dark gray) reveal the
  morphology of the antisymmetric jets, which are emanating from the
  central source.}
  \label{f2}
\end{figure}

On the other hand, from Fig.\ref{f1} and Fig.\ref{f2}, the jet lobes look 
like as cavities of low density and high temperature gas.

In order to make a direct comparison of the numerical results with
the observations, one can generate synthetic emission maps at the
desired wavelength or energy band (the X-ray emission maps will be
discussed later in Section \ref{sec:little}). In the case of radio
emission maps, we could not calculate these because the magnetic
field is not included in our description. Anyhow, it is
possible to use other tracers for the jet structure.  Radio emission
from extragalactic jets is due to synchrotron emission (i.e., of
non-thermal origin), which is enhanced in sites
of strong shock waves. These sites have high temperatures.
Based on this fact, maps of the temperature integrated along
lines-of-sight are expected to be good tracers of the radio-jet
structure. Figure \ref{f2} shows the time evolution of the average
integrated temperature from time $t=8.9$~Myr to $t=130$~Myr.
High-temperature sites reveal the jet morphology. In this
figure we notice an asymmetry between the upper and lower lobe,
which is due to the wind-jet interaction. The characteristic sizes of
these lobes are 10$\times$4~kpc (or 50$\times$20~pixels).

\subsection{Synthetic X-ray emission maps}
\label{sec:little}

\begin{figure}[!t]
\includegraphics[width=\columnwidth]{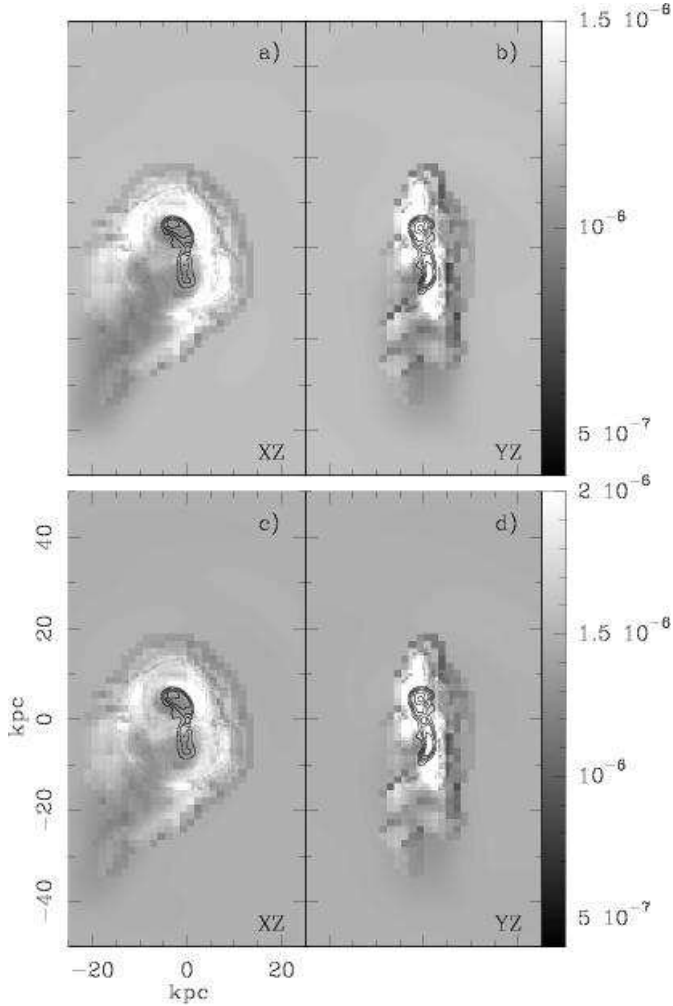} 
\caption{X-ray
emission maps generated from model I. Panels a) and b) show, in
gray-scale, the emission in the soft energy band (0.3--1.8 keV) for the
$xz$ and $yz$ projection, respectively.  The same projections but for
the hard band (1.8--7 keV) are shown in panels c) and d),
respectively.  Each map is overlayed with average temperature contours
in the [2--8]$\times 10^7$~K range. The vertical bars to the right of
panels b) and d) represent the linear gray scale of the X-ray
emission, in units of erg s$^{-1}$ cm$^{-2}$ sr$^{-1}$. All maps
correspond at the same evolution time of 121 Myr. Notice that darker
areas correspond to lower emission regions.}  \label{f3}
\end{figure}

Following the procedure described in subsection \ref{genxray}, we
derive X-ray emission maps by integrating the X-ray emission
coefficient along the selected line-of-sight. In Fig.\ref{f3} we
display the  simulated  X-ray maps for model I for two different
projections on the plane of the sky.  The upper left panel shows
the X-ray map obtained integrating the emission
coefficient along the $y-$axis ($xz$ projection), while
the upper right panel displays the map obtained when the integration
is carried out along the $x-$axis ($yz$ projection). The top two
panels correspond to the X-ray emission in the soft energy band
(0.3--1.8 Kev), for an evolution time of 121 Myr. The same is time-frame is
shown in the bottom two panels (i.e. for the two projections), but for the hard
energy band (1.8--7 Kev) X-ray emission. In order to compare the X-ray
emission with the morphology of the jets, all gray-scale X-ray maps are
overlayed with {\em contours} of the corresponding temperature maps
(averaged along the same lines-of-sight).

From the simulated X-ray maps, we see in both Figs. 3a and 3c (the
$xz$ projection) that the maps show a bright X-ray filament
characterized by an overall bow-shock shape. The wake, observed in the
density distribution maps, presents X-ray emission which is lower than the one
of the intra-cluster medium emission. It is also low the X-ray emission from
the lobe jets.  In contrast, in Figs. 3b and 3d
(the $yz$ projection), the bright bow shock filament is now observed
as a more linear structure superposed on the jets rather than to one side.
This fact shows that the orientation of the flow with respect to the
observer has a strong effect on the predicted 
X-ray emission maps. In other astrophysical 
scenarios, such as in stellar winds and supernova remnants, the importance of 
projection effects was pointed out by Hnatyk \& 
Petruk (1999), Petruk (2001) and Vel\'azquez et al. (2004) . In our case, 
we show that in the $yz$ projection the cavities (at the lobe jet positions) and wake  (observed in the
$xz$ projection) practically 
disappear because they are hidden by the bright emission from the bow shock 
structure, which is located along the same lines of sight.

\begin{figure}[!t]
  \includegraphics[width=\columnwidth]{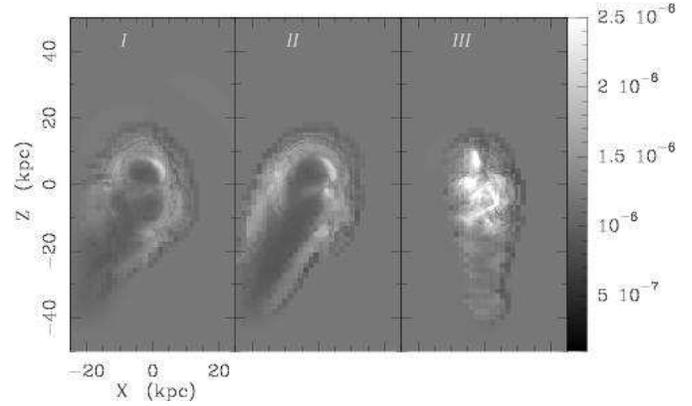}
  \caption{X-ray emission maps in the soft energy band (0.3--1.8~keV),
for models I, II and III (left, middle and right panels,
respectively). The X-ray emission from Model II (with a precession
period five times shorter than model I) looks similar to that of model
I. The X-ray emission from model III is quite different from the other
two, showing several bright, aligned filaments. The three maps
correspond to an integration time of 121 Myr. The linear gray scale of
the X-ray emission is represented by the vertical bar on the
extreme right, in
units of erg s$^{-1}$ cm$^{-2}$ sr$^{-1}$. Notice that darker
areas correspond to lower emission regions.}
  \label{f4}
\end{figure}

We have verified how sensitive model I is with respect to
the adopted parameters.  Fig.\ref{f4} shows the X-ray emission for
models I, II and III, in the soft energy band (0.3-1.8 keV). The X-ray
emissions from models I and II are very similar even though the
precession period is five times faster in model II. However,
in model III the morphology of the X-ray emission changes drastically
as a result of having the plane-parallel wind moving exclusively in
the $z$ direction. The X-ray emission of model III can be
described as being made of several centrally positioned filaments,
surrounded by a lower X-ray emission gas. The morphology shown in
model III looks similar to that observed in radio galaxy jets such
as 3C 294 (Crawford et al. 2003).

\begin{figure}[!t]
  \includegraphics[width=\columnwidth]{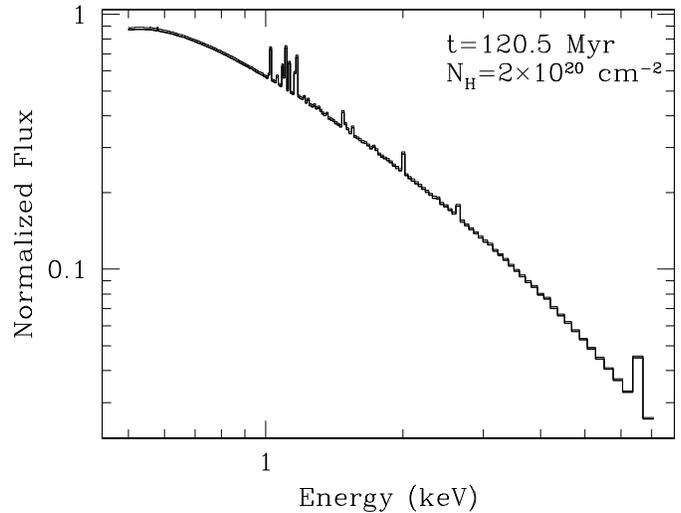}
  \caption{Normalized synthetic spectra from models I, II and III. 
These spectra correspond to an evolution time of 121 Myr and include
absorption by a $N_H=2\times 10^{20}$~cm$^{-2}$ hydrogen column density.}
  \label{f5}
\end{figure}

For each map of models I, II and III, we have derived the
corresponding energy distribution, or synthetic spectrum, in the
 0.3--7.0 keV interval, which are plotted in Fig.\ref{f5}.
These take into account the absorption by an interstellar
absorption corresponding to a $N_H=2\times
10^{20}$~cm$^{-2}$ hydrogen column density.
The three spectra are identical and show spectral
features that were reported by others, such as the presence of some Fe
lines close to 1 keV (see Peterson 2003). 

\section{Analysis of results and comparison with observations}
\subsection{Radio jet in Abell 1795 galaxy cluster}

\begin{figure}[!t]
  \includegraphics[width=\columnwidth]{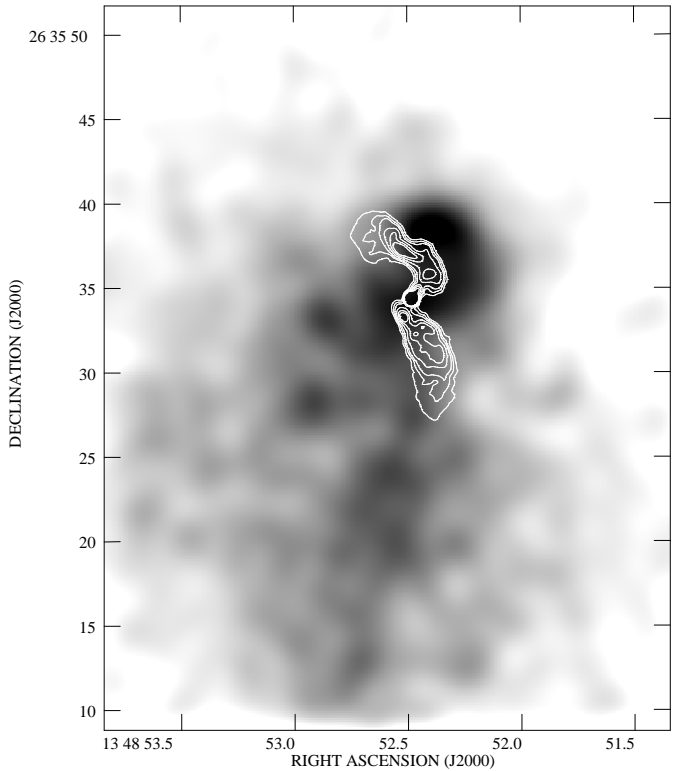}
  \caption{Overlay of the radio-continuum emission at 8.3GHz (in white
    contours, Ge \& Owen 1993) with X-ray diffuse emission from A1795 cluster
    (in greyscale, Fabian et al. 2001). The contours corresponds to the 0.1,
    0.2, 0.3, 0.5 and 1. mJy beam$^{-1}$ levels. Dark areas in greyscale 
correspond to strong enhancement in X-ray emission.}
  \label{f6}
\end{figure}

As a prototype of filamentary structure in X-ray emission, we have the Abell 1795 cluster case. 
Abell 1795 (hereafter A 1795) is a rich cluster of
galaxies, which has been studied at optical, radio and X-ray
wavelengths. Fig.\ref{f6} shows an overlay of the X-ray emission (in greys) from
this cluster (obtained with {\sc Chandra} satellite, Fabian et al. 2001), and
radio-continuum, at 8.3GHz, of the central region (white contours, Ge \& Owen
1993).

Radio observations of the central region of A 1795 show a jet-like
emission from the bipolar radio source 4C 26.42 with a ``S'' shape (see 
white contours in Fig.\ref{f6}). The 
projected combined size
\footnote{we used H$_{0}$=70~km~s$^{-1}$~Mpc$^{-1}$ finding that
$1''$=1.31~kpc.}
is of $\sim 5.6\times 16.2$\, kpc (van Breugel et al. 1984 and Ge \& Owen 1993). 
In X-rays, high resolution images from {\sc Chandra},
show the presence of a long ($\sim 80$~kpc) knotty filament in North-South
direction (see Fig.\ref{f6}). This filament is formed by hot gas, with
 temperatures of the order of 2.7 keV (by
a three ray colour analysis, Fabian et al. 2001).
This large scale filament positionally coincides rather well with an 
elongated optical
H$\alpha$ + [NII] emission line feature previously observed by
Cowie et al. (1983).
Ettori et al. (2002) carried out a similar study by focusing on the
characteristics of the ICM of A 1795. From spectral analyses, they found
that the temperature of the ICM is in the [3.5-4.5]~keV range, being 
higher than the X-ray filament temperature reported by Fabian et al.(2001).

\subsection{Comparison between observations and main numerical results} 

The main results obtained from our numerical simulations are:

\begin{itemize} 
\item by means of integrated temperature maps
(Fig.\ref{f2}), our model describe quite well the shape of the
radio-jets emanating from the central galaxy of A 1795, giving physical
dimensions for the lobes of $\sim 10$~kpc,  which is in
quantitative agreement with the radio image obtained by van Breugel et al.
(1984).
\item the wake left behind by the interaction of the jets with the
plane-parallel wind, as observed in the {\em density} maps
(Fig.\ref{f1}), resembles the morphology of the observed X-ray
filament of A 1795 (Fig.\ref{f6}). However, this ``wake'' cannot be the 
origin for the bright filament in A1795 due to the fact that its X-ray 
emission is lower than the ICM emission.
\item bright X-ray emission filaments have been obtained from our 
three models (specially in the soft X-ray band). 
Simulated X-ray emission maps show  a ``bow shock'' feature (in the
$xz$ projection), which is a consequence of the jet-plane parallel wind 
interaction (see the left panels of Fig. \ref{f3} and panels I and II of 
Fig.\ref{f4}). This ``bow shock'' feature is seen as a long, straight 
filament in the $yz$ 
projection maps (Fig.\ref{f3}). By direct comparison with the observational
results of Fabian et al. (2001) and Ettori et al. (2002), models I and II 
better describe the X-ray emission of A 1795 because the simulated X-ray
emission maps exhibit a single bright filament (compare $yz$~projection of 
Fig.\ref{f3} with Fig.\ref{f6}).
The simulated maps from model III can be employed for explaining the X-ray 
morphology in 3C 294 (Crawford et al. 2003). 
\item the simulated, bright X-ray filaments correspond to emission from gas
with temperatures of a factor $\sim 1.4$
higher than the ICM temperature. This result differs from the
observational result $T_{filament}/T_{ICM}=0.66$ obtained from Fabian et
al. (2001) and Ettori et al. (2002). A discussion about this discrepancy will
be given in the following section.
\item synthetic spectra from three models are very similar to each other, and
 also are in good qualitative agreement with the observations. 

\end{itemize}

\section{Discussion}

We have computed models of a precessing, bipolar jet ejected by a radio galaxy
in motion with respect to the surrounding intra-cluster environment. Our
models have been computed for parameters relevant for the jets and X-ray
filament associated with the 4C 26.42 galaxy in the A 1795 cluster.
Previous works (Reynolds et al. 2001 and Zanni et al. 2003)
 which have studied jet-ICM interaction, predict a shell-like morphology (for
supersonic cocoon expansion regime) or partial shell-like shape (for subsonic
cocoon expansion regime) in X-ray emission. Furthermore the predicted X-ray 
emission is symmetric. The main difference between these models and our work 
is that the jet source is moving into the ICM, 
generating an enhancement in the part of cocoon located in the direction of 
the incoming ICM wind.

From an analysis of our numerical results, we conclude that a precessing 
jet interacting
with a plane-parallel ambient wind qualitatively explains many of the
morphological features observed in A 1795. A jet ejected
from a source moving within the intra-cluster environment with the
velocity of 4C 26.42 leaves behind a wake with a morphology similar to the
one of the X-ray filament associated with this object (Fabian et al. 2001).
The jet precession, at the same time, produces lobes with a 
radio morphology
similar to the one of the radio jets ejected by 4C 26.42 (van Breugel et
al. 1984, Ge \& Owen 1993). Besides, the jet precession has a secondary effect 
producing a
knotty wake structure. To this extent, our numerical models are 
successful at reproducing
the characteristics of the jet+X ray filament system associated with this
galaxy.

However, Ettori et al. (2002) and Fabian et al. (2001) found that the 
X-ray spectrum of the A 1795 filament and its surrounding intra-cluster
medium implies two different temperature. Fabian et al. (2001) found
 the temperature of the
filament might be lower than 2.7~keV, while Ettori et al. 2002 give a
temperature between 3.5 and 4.5~keV. Rodr\'\i guez-Mart\'\i nez et al. 2005
are performing a new analysis using all archived
{\sc Chandra} data of A 1795. 

Our non-radiative models
clearly do not reproduce this effect, as the temperatures obtained
for the gas within the wake are always higher than the temperature
of the surrounding, unperturbed intra-cluster medium.
This failure at obtaining the right temperature is a clear feature
of our present models. However, it might be possible to
reconcile our model with the X-ray observations of A 1795 by including
thermal conduction and/or turbulent mixing of the hot, post-bow shock
gas with material in a cold phase (of $\sim 10^4$~K) belonging {\it in
the past} to the interstellar medium of the central galaxy. In order to
include such effects, much higher resolution simulations would be necessary,
and radiative cooling and thermal/turbulent mixing terms would have to be
included.

Notwithstanding this discrepancy, it is interesting that such a
simple model reproduces the {\it
morphology} of bright X-ray filaments similar to the one observed in A
1795 and other clusters.  Structures with sizes similar to the
observed X-ray filament can be generated in our model. Furthermore,
observed spectral features such as the emission by very high
ionization Fe lines are present in our synthetic spectra.  Of the
three models presented in this work, we consider as acceptable both
models I and II, with model III used only to explore the impact of varying
the incidence angle of the wind.

\label{sec:EPS}

\begin{acknowledgements}

The authors acknowledge support from CONACyT grants 36572-E, 40096-F,
41320-E, 43103-F and DGAPA-UNAM grants IN112602, IN113605 and IN118601.
We thank the anonymous referee for valuable comments that helped 
improve our work. 
We are grateful with Andy C. Fabian and Gregory B. Taylor for the kind
help on provided us their radio and X-rays images of A1795.
Finally, we thank Israel D\'\i az for maintaining
our Linux servers, where all the numerical simulations have been
carried out.
\end{acknowledgements}

\end{document}